\documentstyle[aps,prl,epsfig]{revtex}

\begin{document}

\draft

\title{Intrinsic decoherence in atomic diffraction by standing light wave}

\author{Stefano Mancini and Rodolfo Bonifacio}

\address{
INFM, Dipartimento di Fisica,
Universit\`a di Milano,
Via Celoria 16, I-20133 Milano, Italy
}

\date{\today}

\maketitle

\widetext

\begin{abstract}
We study the effects of time uncertainty in the interaction of atoms
with a standing light wave.
We discuss its physical origin and the possibility to observe
intrinsic decoherence effects by measuring 
the atomic momentum distribution.
\end{abstract}


\section{Introduction}

Typically the decoherence effects are ascribed to the action of an 
environment, i.e. not controlled degrees of freedom.
This is the so called environment-induced decoherence \cite{EID}.
However, there could also be the possibility of an {\it intrinsic} 
nature of decoherence, i.e. coming from the system itself
and not from external degrees of freedom, as pointed out
recently \cite{RBQCM}.
The idea underlying this approach is the fact that the 
interaction time, i.e. the time interval in which the effective 
Hamiltonian evolution takes place, is a random variable \cite{RB}. 
This randomness can have different origins depending on the studied system.

In atom optics \cite{SPECIAL}, for instance,
the interaction time is determined by the 
transit time of the atoms through a cavity \cite{GOULD}.
Nevertheless, this interaction time is random,
due to eventual classical fluctuations of atomic velocities,
but beyond that, due to
the finite extend of the interacting atomic wave packet \cite{MES}.
In fact, the instant at which a wave packet passes a given point
of the axis is not determined precisely, but carries an uncertainty
proportional to its width \cite{MES}. 
Hence, there should be purely quantum mechanical 
intrinsic decoherence effects. 
Here, we study the appearance of such effects in 
atomic diffraction by standing light wave.

On the other hand, decoherence effects relevant to quantum optics domain 
are now beginning 
to be tested experimentally \cite{HAR}. As far as quantum objects
bearing mass are concerned, decoherence has been only 
investigated for the 
motional states of the ions \cite{MONROE}.
Then, it would be also interesting to experimentally verify 
such a new type of decoherence.
To this end, we show a possible modification of the measurable 
atomic momentum distribution.

\section{Time formalism}

In this section we
review the formalism describing non-dissipative decoherence 
derived in Refs. \cite{RB}. It is based
on the idea that time is a random variable or, alternatively, 
that the system Hamiltonian (therefore
its eigenvalues) fluctuates. This leads to random phases in the 
energy eigenstates representation.
Then, the resulting evolution of the system must be averaged on 
a suitable probability 
distribution, and this leads to the decay of the off-diagonal 
elements of the density operator. 

Let us consider an initial state $\rho(0)$ and consider the 
case of a random evolution time. Then,
the evolved state will be averaged over a probability distribution 
$P(t,t')$, i.e.
\begin{equation}\label{rhobardef}
{\overline\rho}(t)=\int^{\infty}_0\, dt' \, P(t,t')\, \rho(t')\,,
\end{equation}
where $\rho(t')=\exp\{-iLt'\}\rho(0)$ is the usual solution of the 
Liouville-Von Neumann equation
with $L\ldots=[H,\ldots]/\hbar$. One can write as well 
\begin{equation}\label{rhobarV}
{\overline\rho}(t)=V_L(t)\rho(0)\,, 
\end{equation} where  
\begin{equation}\label{VP} 
V_L(t)=\int^{\infty}_0\, dt' \, P(t,t')\,e^{-iLt'}\,.  
\end{equation} 
In Ref. \cite{RB}, the
function $P(t,t')$ has been determined to satisfy 
the following conditions:
i) ${\overline\rho}(t)$
must be a density operator, i.e. it must be self-adjoint, 
positive-definite, and with unit-trace.
This leads to the condition that $P(t,t')$ must be 
non-negative and normalized, i.e. a probability
density in $t'$, so that Eq.(\ref{rhobardef}) is a completely 
positive map; ii) $V_L(t)$ satisfies
the semigroup property $V_L(t_1+t_2)=V_L(t_1)V_L(t_2)$, 
with $t_1, t_2 \ge 0$. These requirements
are satisfied by 
\begin{equation}\label{V} V_L(t)=\frac{1}{(1+iL\tau)^{t/\tau}}\,,  
\end{equation} 
and  
\begin{equation}\label{P} 
P(t,t')=\frac{1}{\tau}\frac{e^{-t'/\tau}}{\Gamma(t/\tau)}
\left(\frac{t'}{\tau}\right)^{(t/\tau)-1}\,, 
\end{equation} 
where the parameter $\tau$ naturally appears as a scaling time. 
The expression
(\ref{P}) is the so-called $\Gamma$-distribution function, 
well known in line theory. Its
interpretation is  particularly simple 
when $t/\tau=r$ with $r$ integer; 
in that case $P(r,t')$
gives the probability density that the waiting time for 
$r$ independent events is $t'$ and $\tau$ is
the mean time interval between two events. Generally, 
the meaning of the parameter $\tau$ can be
understood by considering the mean of the evolution 
time $\langle t'\rangle=t$, and its variance
$\langle t'^2\rangle-\langle t'\rangle^2=\tau t$.  

When $\tau\to 0$, $P(t,t')\to \delta(t-t')$ so
that ${\overline\rho}(t)\equiv\rho(t)$ and $V_L(t)=\exp\{-iLt\}$ 
is the usual unitary evolution.
However, for finite $\tau$, the evolution operator $V_L(t)$ 
describes a decay of the off diagonal
matrix elements in the energy representation, whereas the 
diagonal matrix elements remain constants,
i.e. the energy is still a constant of motion. In fact, 
in the energy eigenbasis,
Eqs.(\ref{rhobarV}) and (\ref{V}) yield  
\begin{equation}\label{rhobarnm}
{\overline\rho}_{n,m}(t)=e^{-\gamma_{n,m}t}\,
e^{-i\nu_{n,m}t}\,\rho_{n,m}(0)\,, 
\end{equation}
where 
\begin{eqnarray}
\gamma_{n,m}&=&\frac{1}{2\tau}
\log\left(1+\omega_{n,m}^2\tau\right)\,,\label{ga}
\\
\nu_{n,m}&=&\frac{1}{\tau}
\arctan\left(\omega_{n,m}\tau\right)\,,\label{nu}
\end{eqnarray} 
with
$\hbar\omega_{n,m}=(E_n-E_m)$ the energy difference. 
One can recognize in Eq.(\ref{rhobarnm}),
beside the exponential decay, a frequency shift of 
every oscillating term. 

The phase diffusion
aspect of the present approach can also be seen in the 
evolution equation for the averaged density
matrix ${\overline\rho}(t)$. Indeed, by differentiating 
with respect to time Eq.(\ref{rhobarV}) and
using (\ref{V}) one gets the following master equation 
for ${\overline\rho}(t)$
\begin{equation}\label{meq} 
\frac{d}{dt}{\overline\rho}(t)= -\frac{1}{\tau}\log
\left(1+iL\tau\right) {\overline\rho}(t)\,. 
\end{equation} 
It is worth noting that by expanding
the logarithm up to second order in $\tau$, one obtains 
\begin{equation}\label{meqapp}
\frac{d}{dt}{\overline\rho}(t)=
-\frac{i}{\hbar}\left[H,{\overline\rho}(t)\right]
-\frac{\tau}{2\hbar^2}\left[H,\left[H,{\overline\rho}(t)
\right]\right]\,,
\end{equation}
which is the well known phase-destroying 
master equation \cite{QO}.
Hence, Eq.(\ref{meq}) appears as a generalized 
phase destroying master equation taking into account
higher order terms in $\tau$. Nonetheless, 
the present approach is different from the
usual master equation approach, in the sense 
that it is model independent, non perturbative 
and without specific statistical assumptions.
In fact, the probability distribution 
(\ref{P}), is derived only from the semigroup
condition, and it is interesting to note that this 
condition yields a gaussian probability
distribution  as a limiting case when $t\gg\tau$.

\section{Atomic diffraction by standing light wave}

\begin{figure}[t]
\centerline{\epsfig{figure=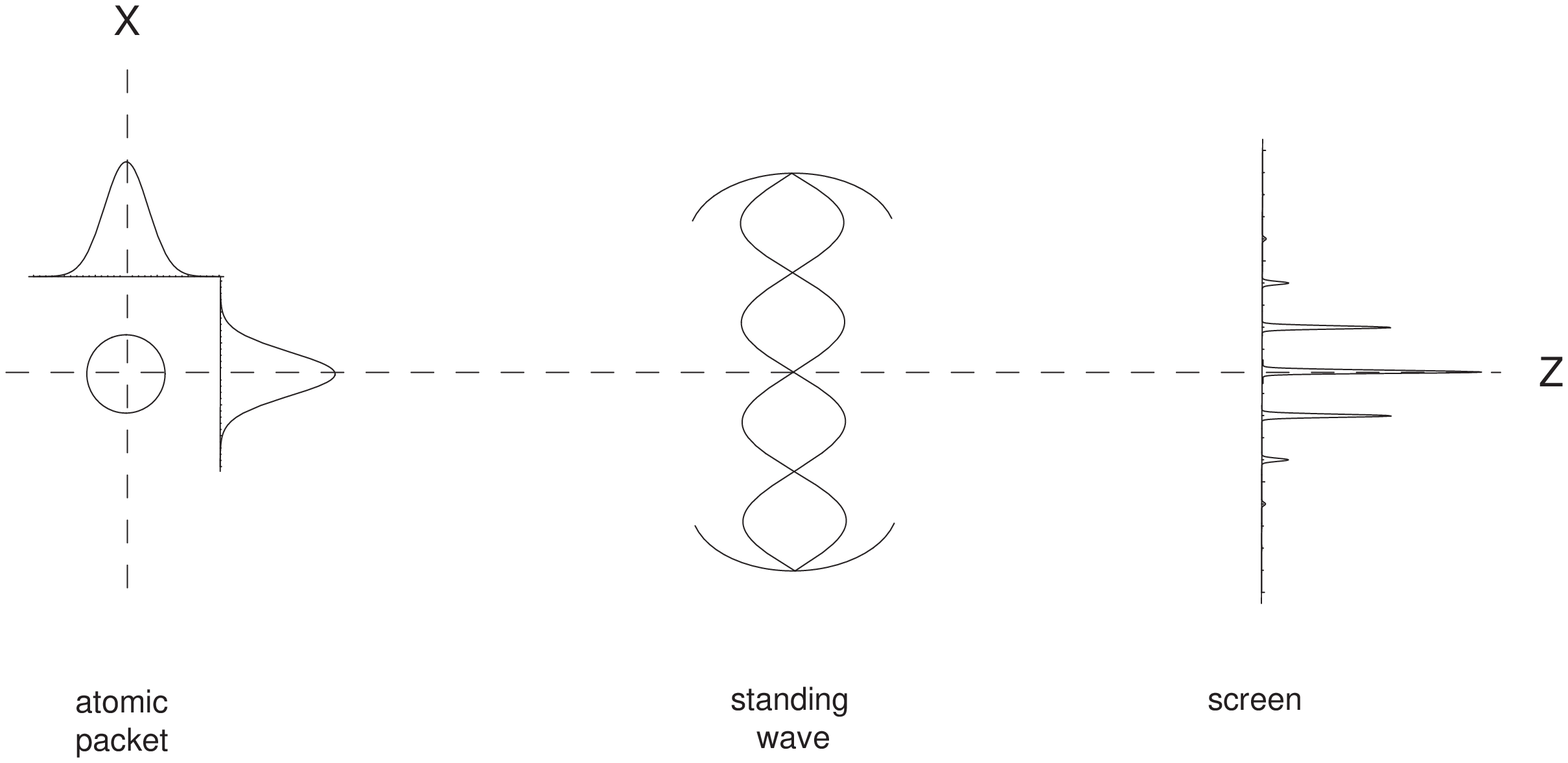,width=5.5in}}
\caption{\widetext 
Scheme
}
\label{fig1}
\end{figure}

We consider the deflection of an atomic beam of two-level atoms 
by a classical standing wave in a cavity as in Fig.(\ref{fig1}).
The atomic beam is normally incident on the standing wave and experiences
an exchange of momentum with the photons in the light wave.
We shall assume that the frequency of the light field is well detuned from 
the atomic resonance so that we may neglect the spontaneous emission.
The Hamiltonian describing the system is \cite{QO}
\begin{equation}\label{Hini}
H=\hbar\omega_0\sigma_Z+\frac{{\bf P}^2}{2M}
+\hbar \Omega\left(\sigma_- e^{-i\omega t}+\sigma_+ e^{i\omega t}\right)
\cos(kX)\,,
\end{equation}
where ${\bf P}\equiv (P_X,P_Y,P_Z)$ is the center of mass momentum 
of the atom, $M$ is the atomic mass, $\sigma_Z$ and $\sigma_{\pm}$ are the 
pseudo spin operators for the atom, $\omega_0$ and $\omega$ are the atomic 
and field frequencies, $k=\omega/c$ is the wave number of the standing 
wave, and $\Omega$ the Rabi frequency.

We shall assume that the interaction time is sufficiently small that the 
transverse kinetic energy absorbed by the atom during the interaction can 
be neglected (Raman-Nath approximation). This is equivalent to neglect
the term $P^2/2M$ in the Hamiltonian (\ref{Hini}).
Then, transforming to the interaction picture (with 
$H_0=\hbar\omega\sigma_Z$) the 
Hamiltonian may be written in the form
\begin{eqnarray}
H&=&\hbar\Delta\sigma_Z+2\hbar\Omega\sigma_X \cos(kX)\,,
\\
&=&\hbar V(X)\left[\cos \theta(X) \sigma_Z+\sin \theta(X) \sigma_X\right]\,,
\end{eqnarray}
where $\Delta=\omega_0-\omega$, 
$V(X)=\sqrt{\Delta^2+4\Omega^2\cos^2(kX)}$,
$\cos\theta(X)=\Delta/V(X)$ and $\sin\theta(X)=2\Omega\cos(kX)/V(X)$.

In the limit of large detuning, i.e. $\Delta \gg 2\Omega\cos(kX)$,
the effective Hamiltonian remains
\begin{equation}\label{Heff}
H=\hbar\Delta\sigma_Z+\left[\frac{2\hbar\Omega^2\cos^2(kX)}{\Delta}
\right]\sigma_Z\,.
\end{equation}

The atomic state vector can be written as 
\begin{equation}
\langle X | \Psi(t) \rangle=a(X,t) |e\rangle
+ b(X,t) |g\rangle\,,
\end{equation}
where $|g\rangle$, and $|e\rangle$ are the ground and excited internal 
atomic states, while $a$, and $b$ are probability amplitudes.

We assume the atoms initially in their ground state with a Gaussian 
wavefunction
\begin{eqnarray}
a(x,0)&=&0\,,
\\
b(x,0)&=&(\pi\sigma^2)^{-1/4} \exp\left[-\frac{x^2}{2\epsilon_x}\right]\,,
\end{eqnarray}
where $x=kX$ and $\epsilon_x$ is the transverse position spread of the input 
beam.

The Schr\"odinger equations are
\begin{eqnarray}
\frac{d}{dt}a&=&i\left[\frac{\Delta}{2}+\frac{\Omega^2\cos^2 
x}{\Delta}\right]a\,,
\\
\frac{d}{dt}b&=&-i\left[\frac{\Delta}{2}+\frac{\Omega^2\cos^2 
x}{\Delta}\right]b\,,
\end{eqnarray}
Of course, it results $a(x,t)=0$, 
while the solution for $b$ can be written in terms of the
Bessel functions of the first kind $J_n$ \cite{MF}, as
\begin{equation}\label{b}
b(x,t)=\exp\left[i\left(\frac{\Delta}{2}+\frac{\Omega^2}{2\Delta}
\right)t\right]\sum_{n=-\infty}^{+\infty} i^n
J_n\left(\frac{\Omega^2 t}{2\Delta}\right) e^{2inx} \, b(x,0)\,.
\end{equation}
Taking the Fourier transform of this relationship shows the effect in the 
momentum space
\begin{equation}\label{btilde}
{\tilde b}(p_x,t)=\exp\left[i\left(\frac{\Delta}{2}+\frac{\Omega^2}{2\Delta}
\right)t\right]\sum_{n=-\infty}^{+\infty} i^n
J_n\left(\frac{\Omega^2 t}{2\Delta}\right) \, {\tilde b}(p_x-2n,0)\,,
\end{equation}
where ${\tilde b}$ denotes the Fourier transform of $b$,
and $p_x=P_X/\hbar k$.
The scattered ground state wavefunction is a superposition of Gaussian 
modulated plane waves with momentum $2n\hbar k$. The momentum transferred 
from the field to the atom is an even multiplies of $\hbar k$ 
corresponding to the absorption of a photon from the $(+k)$ component 
followed by induced emission in the $(-k)$ component of the standing wave.

The final output momentum probability, which can be measured, is composed of a 
comb of images of the initial momentum distribution
\begin{equation}\label{wp}
w(p_x,T)=\left| {\tilde b}(p_x,T) \right|^2
=\sum_{n=-\infty}^{+\infty} \sum_{m=-\infty}^{+\infty} i^{(n-m)}
J_n\left(T/4\right) J_m\left(T/4\right) \, 
{\tilde b}(p_x-2n,0) \, {\tilde b}(p_x-2m,0)
\,,
\end{equation}
where $T=2\Omega^2 t/\Delta$.
In order to resolve the peaks it is necessary to have a narrow initial 
momentum spread, i.e. $\epsilon_x > 1$.

\section{Effects of time fluctuations}

The atomic wave packet has a finite position spread also 
in the longitudinal direction ($Z$ in Fig.1). 
For the sake of simplicity we assume 
$\epsilon_z=\epsilon_x=\epsilon=k\epsilon_Z=k\epsilon_X$.
The non-zero value of $\epsilon_z$ give rise to
an uncertainty in the arrival time \cite{MES} of the atoms
at the end of the cavity, i.e. at the time $T$ of Eq.(\ref{wp}).
We can account for this uncertainty by considering 
the time as a random variable and use the 
quantum mechanical consistent approach described in
Section II \cite{AHA}.
Then, accordingly to Eq.(\ref{rhobardef}), 
we have to do the following replacement
\begin{equation}
w(p_x,T)\equiv{\rm Tr}\left\{p_x\,\rho(T)\right\}
\longrightarrow
\overline{w}(p_x,T)\equiv{\rm Tr}\left\{p_x\,\overline{\rho}(T)\right\}
\,.
\end{equation}
From Eqs.(\ref{btilde}) and (\ref{P}), it results
\begin{equation}
\overline{w}(p_x,T)=\sum_{n=-\infty}^{+\infty} \sum_{m=-\infty}^{+\infty} 
i^{(n-m)}
I_{n,m}(T) \, 
{\tilde b}(p_x-2n,0) \, {\tilde b}(p_x-2m,0)
\,,
\end{equation}
where 
\begin{equation}\label{Inm}
I_{n,m}(T)=\int_0^{\infty} \, dT' \,
J_n(T'/4)\,J_m(T'/4)\,
\frac{1}{\cal T}\,\frac{e^{-T'/{\cal T}}}{\Gamma(T/{\cal T})}\,
(T'/{\cal T})^{T/{\cal T}-1}\,,
\end{equation}
with ${\cal T}=2\Omega^2\tau/\Delta$,
and the integral given explicitly in the appendix.

The output momentum distribution is shown in Fig.2 for several values of 
${\cal T}$. Notice, that the chosen interaction time $T$ 
implies the deflection of all atoms in the ideal situation a), where 
no peak is present at $p_x=0$.
However, the effect of intrinsic decoherence is to reduce the exchange of 
momentum between atoms and field, then to leave the atoms undeflected.
In fact the central peak increases from b) to c) while the others decrease. 

\begin{figure}[t]
\centerline{\epsfig{figure=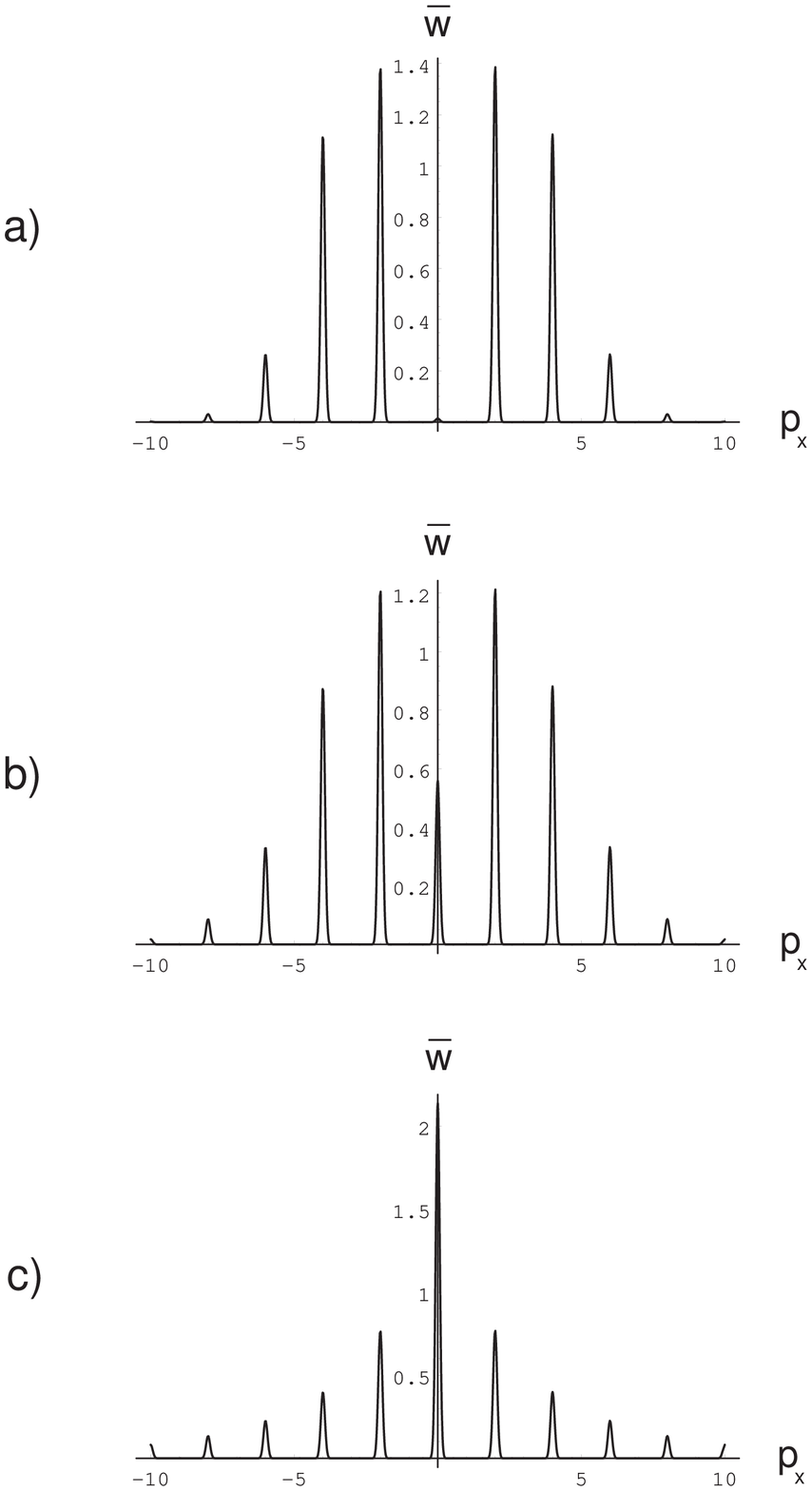,width=3.0in}}
\caption{\widetext 
Momentum distribution $\overline{w}$ for 
${\cal T}=0$ a); ${\cal T}=1$ b); and ${\cal T}=10$ c).
The values of other parameters are: $T=10$, and $\epsilon=10$.
}
\label{fig2}
\end{figure}

\section{Conclusions}

We know that ${\cal T}$ is related to the time parameter $\tau$.
The latter represents the time uncertainty in the 
interaction process.
On the other hand, the uncertainty in localizing temporally a 
wave packet at a given space point, which give rise to the
time energy uncertainty relation \cite{MES},
may be considered as
$\delta Z/v$, where $\delta Z$ is the width of the packet and $v$
its group velocity \cite{MES}. 
Then, we may assume \cite{RBQCM}
\begin{equation}\label{tau}
\tau=\frac{1}{\langle P_Z\rangle/M}
\,\sqrt{
\epsilon_Z^2+\left(\frac{\hbar}{2M}\right)^2
\frac{t^2}{\epsilon_Z^2}
+\left(\frac{1}{M}\right)^2{\cal E}_{P_Z}^2t^2
}
\,,
\end{equation}
where $\langle P_Z \rangle/M$ is the mean velocity of the atoms along the $Z$ 
direction, while the square root term represents the width of the packet
along the $Z$ direction.
The first two terms inside the square root of 
Eq.(\ref{tau}) describe the well known free particle wave packet spread
(during the interaction time $t$), 
while the last term is due to the classical momentum spread.
In this case $\tau$ (hence ${\cal T}$) becomes time
dependent, but the positive map described by Eqs.(\ref{rhobardef}), 
(\ref{P}), can be maintained by dropping the semigroup property.

Of course, if in the above expression (\ref{tau}) 
${\cal E}_{P_Z}$ is the dominant term,
we only have classical decoherence effect, essentially due to the 
thermal spread of momentum.
However, even if such effect is eliminated, decoherence will be still 
present having a purely quantum nature.
Moreover, it will be intrinsic to the system, and not due to external 
degrees of freedom.

Effects similar to those of Fig.(\ref{fig2}) 
can be recognized in the experimental data of Ref.\cite{GOULD}, 
but mainly due to (classical) velocity spread. 
Nevertheless, more recently, 
cooling techniques offer the possibility to have 
almost monochromatic atomic beam 
(e.g. $\delta \langle P_Z \rangle/ \langle P_Z \rangle=10^{-3}$ with
$\langle P_Z \rangle /M=10^3$ ${\rm m s}^{-1}$ \cite{FAU}).
Then, it would be possible to achieve 
minimum uncertainty states for the atomic wave packet, with
e.g. $\epsilon_Z\approx 10^{-11}$ 
${\rm m}$ \cite{FAU}. 
This means that the dominant term in Eq.(\ref{tau}),
once fixed the interaction time $t\approx 10^{-9}$ ${\rm s}$ 
\cite{GOULD}, becomes the Schrodinger 
spread, i.e. the second term in the square root.
That gives ${\cal T}\approx 1$ 
which is promising to test the intrinsic quantum nature of decoherence
as can be seen from Fig.(\ref{fig2}).
On the other hand, the opposite limit, where $\epsilon_Z$ is dominant 
in Eq.(\ref{tau}) could be investigated as well with slow atoms at 
large de Broglie wavelength \cite{CHU}.

In conclusion, we have studied intrinsic decoherence 
arising from the interaction time fluctuations in 
atomic diffraction by 
standing light wave. 
The main limitation to observe such effect relies on the necessity
to suppress classical 
fluctuations, however, the technological achievements in atom optics,
atom lasers \cite{BOSER} at last, make the possibility to observe 
{\it intrinsic quantum decoherence} realistic.
Finally, it is worth interesting to also apply the above theory
to the problem of atomic wave diffraction in time \cite{MOSH};
this is planned for future work.

\section*{Appendix}

The integral (\ref{Inm}) can be written explicitly \cite{RYZ}
as follows
\begin{equation}
I_{n,m}(T)=I_{|n|,|m|}(T)\,(-)^{(|n|-n+|m|-m)/2}\,,
\end{equation}
where
\begin{equation}
I_{|n|,|m|}(T)=
\frac{
{\cal T}^{(|n|+|m|)}
\Gamma\left(|n|+|m|+\frac{T}{\cal T}\right)
{}_4F_3\left({\bf u},{\bf v},-\frac{{\cal T}^2}{4}\right)}
{2^{3(|n|+|m|)}
\Gamma(1+|n|) \Gamma(1+|m|) \Gamma\left(\frac{T}{\cal T}\right)}
\,,
\end{equation}
with $F$ the generalized hypergeometric function \cite{MF} 
whose vectors ${\bf u}$ and ${\bf v}$ are
\begin{eqnarray}
{\bf u}&\equiv&\left(
\frac{1}{2}+\frac{|n|}{2}+\frac{|m|}{2}\,,
\frac{1}{2}+\frac{|n|}{2}+\frac{|m|}{2}\,,
\frac{|n|}{2}+\frac{|m|}{2}+\frac{T}{2{\cal T}}\,,
\frac{1}{2}+\frac{|n|}{2}+\frac{|m|}{2}+\frac{T}{2{\cal T}}\right)\,,
\\
{\bf v}&\equiv&\left(1+|n|,1+|m|,1+|n|+|m|\right)\,.
\end{eqnarray}

\end{document}